\input{aipcheck}

\documentclass[
    ,final            % use final for the camera ready runs
  ]
  {aipproc}

\layoutstyle{6x9}

\begin{document}

\title{Strong spectral evolution during the prompt emission of GRB\,070616}


\author{R.L.C. Starling}{
  address={Dept. of Physics and Astronomy, University of Leicester, University Road, Leicester, LE1 7RH, UK}
}

%\author{<author2>}{
%  address={<common address for author2 and author3>}
%}

%\author{<author3>}{
%  address={<common address for author2 and author3>}
%  ,altaddress={<author1 address>} % additional visiting address
%}

\begin{abstract}
{\it Swift} has revealed features in GRB early light curves, such as steep decays and
X-ray flares, whose properties are consistent with an internal origin though
they are far from understood. The steep X-ray decay is often explained using
the curvature effect; however a significant number of GRBs display strong
spectral evolution during this phase, and a new mechanism must be invoked to
explain this.
Of particular interest are the longest duration GRBs in which the early emission can be
studied in most detail. Here we present data for GRB\,070616, in which the prompt
emission shows a complex multipeaked structure, leading to one of the
longest prompt emission durations ever recorded. We take advantage of extensive coverage of such a long
burst by all Swift instruments. Combining data from {\it Swift} and {\it Suzaku} we study
the evolution of the prompt emission spectrum, following the temporal
variability of the peak energy and spectral slope.

\end{abstract}

\maketitle

%%%%%%%%%%%%%%%%%%%%%%%%%%%%%%%%%%%%%%%%%%%%
%% MAINMATTER
%%%%%%%%%%%%%%%%%%%%%%%%%%%%%%%%%%%%%%%%%%%%
 
\section{Introduction}
The prompt emission mechanism for Gamma-ray Bursts (GRBs) is usually
attributed to internal shocks due to collisions of shells of different
Lorentz factors ejected from the vicinity of a compact object
\citep{Rees}. Though the general picture appears applicable to most GRBs, the
details are far from understood and other models have also been proposed
\citep[see][and references therein]{Zhang1}. The {\it Swift}
mission
%\citep{Gehrels1} 
has revealed complex early emission, as yet largely unexplained. The steeply
decaying phases that directly follow both prompt emission and X-ray flares are
often interpreted as due to the curvature effect \citep[e.g.][]{Kumar2}
where high latitude emission is delayed with respect to that on-axis. However,
in this scenario significant spectral evolution is not expected but is seen in
a large number of cases \citep{Zhang2}.
The longest duration GRBs provide rare opportunities to study the
relationship between various possible early emission components. Very few GRBs are detected in $\gamma$-rays
for more than 400 seconds when using the T$_{\rm 90}$ parameter. Only
$\sim$0.5\% of the BATSE sample meet this criteria \citep{Paciesas} and such GRBs
remain rare in the {\it Swift} era.

Here we present {\it Swift} and {\it Suzaku} data for GRB\,070616, whose T$_{90} =
402\pm10$ s is one of the longest on record. We
study the evolution of the prompt
emission through broadband lightcurves and spectra. Further details of this
study can be found in \cite{Starling}.

\section{Results}
%\subsubsection{Temporal analysis}
{\it Swift} BAT
triggered on a gradual rise of $\gamma$-ray emission which lasted approximately 100 s before the
first and strongest peak, centred at T$_{\rm 0}$+120 s. Thereafter multiple blended peaks continue
the seemingly flat prompt emission in both $\gamma$-rays and X-rays out to T$_{\rm 0}$+500--600 s. At this point the $\gamma$-ray
emission appears to return to the count rate at which it began at T$_{\rm 0}$, whilst the X-ray
emission begins a very steep decay lasting until T$_{\rm 0}$+1200 s. Fig. 1 shows
the joint BAT-XRT lightcurve in which peaks are temporally coincident strongly suggesting that the X-ray and $\gamma$-ray
emission come from the same component. Contrastingly,
the $V$ band optical emission observed with UVOT is rising.

The BAT hardness ratio (50--100 keV/15--50 keV) remained approximately constant
until T$_{\rm 0}$+285 s when the spectra softened
significantly over the remainder of the $\gamma$-ray observations. The XRT hardness
ratio (1--10 keV/0.3--1 keV) shows the same behaviour, with the spectral
evolution beginning at $\sim$T$_{\rm 0}$+500 s.
%\subsubsection{Time dependent spectral analysis}
To examine the spectral evolution seen in the hardness ratios, we time-sliced
the BAT and XRT data into 100s-long bins covering T$_{\rm 0}$+137--737 s. 
We adopted an absorbed Band function model \citep{Band}, fixing the high energy power law slope to $\beta=2.36$ as found in the
{\it Suzaku} spectrum \citep[][and adopting $F_{\nu} \propto
\nu^{-\beta}$]{Morigami,Starling}. Intrinsic extinction was determined from fits to the
X-ray data alone to be constant, amounting to a total Galactic+intrinsic
absorbing column of $N_{\rm H} = 0.4 \times 10^{22}$ cm$^{-2}$ (no redshift is
available). The peak
energy is derived from the free parameters $\alpha$ (low energy power law
index) and $E_{\rm 0}$ (characteristic energy) using $E_{\rm pk} = E_{\rm 0} (2-\alpha)$.
%\begin{equation}
%E_{\rm pk} = E_{\rm 0} (2-\alpha).
%\end{equation}
$E_{\rm pk}$
can be well constrained and is observed to move to lower
energies with time from 135 keV down to 4 keV in $\sim$600 s, while the spectral slope
$\alpha$ also varies gradually, softening with time, shown in Fig. 2.

\section{Discussion and conclusions}
The lightcurve is atypical in that the
emission rises relatively slowly over $\sim$100 s to a peak,
then persists at a fairly constant level before
showing a rapid decline. Strong spectral evolution is observed
throughout the prompt phase. We are able to track the spectral peak energy as
it moves from the WAM/BAT energies down to a few keV. At the same time we find
a softening of the spectrum below the peak energy.

Spectral evolution through the prompt phase has been noted previously, and is inconsistent with the idea that
the curvature effect alone is driving the
emission during the steep decay \citep[e.g.][]{Liang}. Using a curvature
effect model to explain the steep decay in GRB\,070616 we require that phase to begin at T$_{\rm 0}$+632$^{+11}_{-12}$ s requiring a long initial emission period
of $>$600 s.
The curvature effect is in fact a poor fit, but the combination of the curvature
effect and the strong spectral evolution
we observe
may be able to account for the steep X-ray decay.

Only 25\% of {\it Swift} GRB
prompt X-ray tails can be
fit with the curvature effect alone \citep{Zhang2}; those not well fit show spectral evolution. From fits to 16 GRBs, \cite{Zhang2} tested and subsequently disfavoured
two possible causes of the spectrally evolving X-ray tails, namely an
angle-dependent spectral index in structured jets and a superposition of the
curvature effect and a power law decay component.
The observed spectral softening could, they suggest, be caused by cooling of the
plasma where the cooling frequency decreases with time. This manifests itself as a cut-off power law shape with the
cut-off moving to lower energies with time, proposed for GRBs 060218 and 980923 and similar to our Band function results for GRB\,070616. In this
scenario the peak energy we track through the BAT band would be the
cooling frequency. In addition we observe softening of the low energy power
law slope which remains unexplained.

It is possible that we are observing a combination of components which
together mimick spectral evolution. We investigated a scenario with two
spectrally invariant power laws: the relative contribution of the softer power
law to the total spectrum increasing with time.
This is a viable explanation of the spectral
evolution, but can be ruled out as it does not fit the lightcurve behaviour. 
We then applied the modelling procedure of \cite{O'Brien,Willingale} in which it was shown
that many GRB early lightcurves can be well modelled by up to two emission components
each consisting of an exponential$+$power law decay. But
again the prompt emission lightcurve of GRB\,070616 is not well fit by this model.
Interestingly, several, although not all,
of the GRBs with strong spectral evolution studied in \cite{Zhang2} are also
not well fitted by the two-component lightcurve model
\citep[e.g. GRBs 051227 and 060614, see][]{Willingale}. Others not
well fitted by this type of modelling (e.g. GRB\,051117A) were
identified by \cite{Zhang2} as having evolution which
they attributed to flares. Such GRBs may instead be more similar to
GRB\,070616.

In conclusion, the prompt emission of GRB\,070616 comprises a component well fitted with a
Band function and a possible further component. The movement of the peak
energy shows that great care must be taken when using average $E_{\rm pk}$ values.
It is clear that both broadband coverage and good time resolution are crucial to
pinning down the origins of the complex prompt emission in GRBs. 

%\subsubsection{<A subsubsection>}
%Some url test \url{http://www.world.universe}.
%
%Et iam nox umida caelo praecipitat $J_{ion}$ suadentque cadentia
%sidera somnos. Sed si tantus amor casus cognoscere nostros et breviter
%Troiae $J_{ion}$ supremum audire laborem:
%\begin{equation}
%J_{ion}=A\frac{exp\left[-\frac{E_a}{kT}\right]}{kT}\alpha \label{ionflux}
%\end{equation}
%lamentabile regnum cruerint Danai; quaeque ipse miserrima vidi, et
%quorum pars magna fui. $A$ talia fando, $E_a$ iam nox umida, $k$ caelo
%praecipitat, suadentque cadentia sidera somnos. See \eqref{ionflux}.

%\paragraph{<A subsubsubsection>}%
%
%Infandum, regina, iubes renovare dolorem, Troianas ut opes et
%lamentabile regnum cruerint Danai; quaeque ipse miserrima vidi, et
%quorum pars magna fui \cite{Brown2000,BrownAustin:2000}. Quis talia fando
%Myrmidonum Dolopumve aut duri miles Ulixi temperet
%\cite{Mittelbach/Schoepf:1990} a lacrimis? Et iam
%nox umida caelo praecipitat, suadentque \cite{Wang} cadentia
%sidera somnos.

%%%%%%%%%%%%%%%%%%%%%%%%%%%%%%%%%%%%%%%%%%%%
%% Sample figure:
%%
%% The option [height=...] scales the picture to the given height,
%% without it it would be printed at its nominal size
%%%%%%%%%%%%%%%%%%%%%%%%%%%%%%%%%%%%%%%%%%%%
\begin{figure}
  \includegraphics[height=.4\textheight,angle=-90]{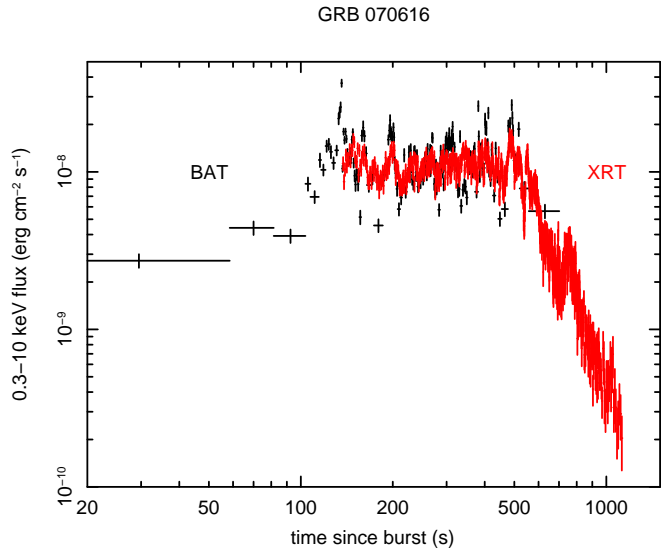}
  \caption{The joint BAT-XRT lightcurve (BAT extrapolated to the XRT band
    using multiple spectral fits) showing the coincidence of peaks, long-lived
    flat behviour and steep X-ray decline.}
\end{figure}

\begin{figure}
  \includegraphics[height=.4\textheight,angle=90]{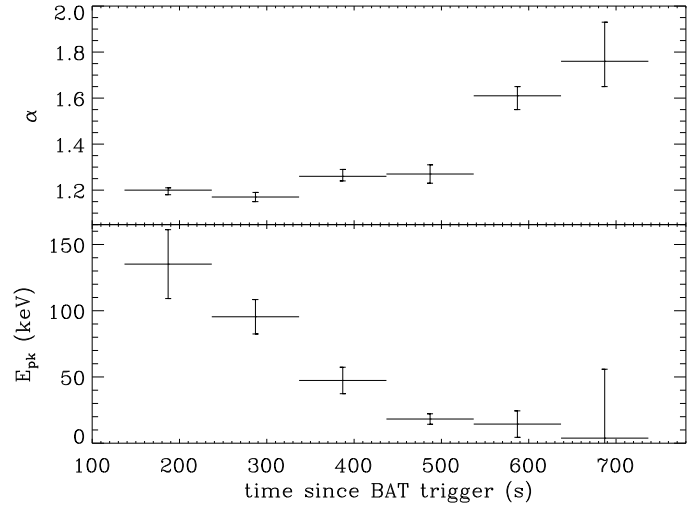}
  \caption{The variability of spectral slope $\alpha$ (upper panel) and peak energy
    $E_{\rm pk}$ (lower panel) in
    Band function fits to the early BAT-XRT spectra.}
  \label{Epkband}
\end{figure}

\begin{theacknowledgments}
We thank the {\it Swift} and {\it Suzaku} WAM teams for their contributions,
in particular P.T. O'Brien, R. Willingale, K.L. Page, J.P. Osborne, M. De
Pasquale, N.P.M. Kuin and M. Tashiro. RLCS acknowledges support from STFC.
\end{theacknowledgments}



\bibliographystyle{aipproc}   % if natbib is available

\bibliography{ms}

\IfFileExists{\jobname.bbl}{}
 {\typeout{}
  \typeout{******************************************}
  \typeout{** Please run "bibtex \jobname" to optain}
  \typeout{** the bibliography and then re-run LaTeX}
  \typeout{** twice to fix the references!}
  \typeout{******************************************}
  \typeout{}
 }

\end{document}